\documentclass[10pt, a4paper]{article}
\usepackage{lrec2022} 
\usepackage{multibib}
\newcites{languageresource}{Language Resources}
\usepackage{graphicx}
\usepackage{tabularx}
\usepackage{soul}

\usepackage[normalem]{ulem}
\useunder{\uline}{\ul}{}

\usepackage{titlesec}
\usepackage{url}
\titleformat{\section}{\normalfont\large\bfseries\center}{\thesection.}{1em}{}
\titleformat{\subsection}{\normalfont\SmallTitleFont\bfseries\raggedright}{\thesubsection.}{1em}{}
\titleformat{\subsubsection}{\normalfont\normalsize\bfseries\raggedright}{\thesubsubsection.}{1em}{}
\renewcommand\thesection{\arabic{section}}
\renewcommand\thesubsection{\thesection.\arabic{subsection}}
\renewcommand\thesubsubsection{\thesubsection.\arabic{subsubsection}}

\usepackage{epstopdf}
\usepackage[utf8]{inputenc}

\usepackage[hyperfootnotes=false]{hyperref}
\usepackage{xstring}

\usepackage{color}
\usepackage{fdsymbol}

\newcommand{\footnoteurl}[1]{\hbadness=10000{\scriptsize\url{#1}}}

\newcommand\blfootnote[1]{%
  \begingroup
  \renewcommand\thefootnote{}\footnote{#1}%
  \addtocounter{footnote}{-1}%
  \endgroup
}

\usepackage[bottom]{footmisc}

\usepackage[inline]{enumitem}

\usepackage{booktabs}
\usepackage{calc}
\usepackage{array}
\newcolumntype{C}{>{\centering\arraybackslash}p{\widthof{\bf RKPQA}}}

\DeclareGraphicsExtensions{.pdf,.ai,.jpg,.png}
\setkeys{Gin}{pagebox=artbox}
\graphicspath{{figures/}}

\begin{document}

\title{SCAI-QReCC Shared Task on Conversational Question Answering}

\name{Svitlana Vakulenko\textsuperscript{$\medstar\spadesuit$}, Johannes Kiesel\textsuperscript{$\vardiamondsuit$}, Maik Fr\"{o}be\textsuperscript{$\varheartsuit$}} 
\address{\textsuperscript{$\medstar$}Amazon, Spain, svvakul@amazon.com\\
         \textsuperscript{$\vardiamondsuit$}Bauhaus-Universität Weimar, Germany, johannes.kiesel@uni-weimar.de\\
         \textsuperscript{$\varheartsuit$}Martin-Luther-Universität Halle-Wittenberg, Germany, maik.froebe@informatik.uni-halle.de\\}

\abstract{Search-Oriented Conversational AI (SCAI) is an established venue that regularly puts a spotlight upon the recent work advancing the field of conversational search.
SCAI'21 was organised as an independent on-line event and featured a shared task on conversational question answering.
Since all of the participant teams experimented with answer generation models for this task, we identified evaluation of answer correctness in this settings as the major challenge and a current research gap.
Alongside the automatic evaluation, we conducted two crowdsourcing experiments to collect annotations for answer plausibility and faithfulness.
As a result of this shared task, the original conversational QA dataset used for evaluation was further extended with alternative correct answers produced by the participant systems.
 \\ \newline  }

\maketitleabstract

\section{Introduction}
\label{introduction}
\blfootnote{$\spadesuit$ Research conducted when the author was at the University of Amsterdam.}

Conversational Question Answering (QA) is a challenging task at the current research frontier~\cite{DBLP:conf/aaai/QiuHCJQ0HZ21,DBLP:conf/acl/KimKPK20} important for developing conversational information retrieval (conversational search) systems~\cite{DBLP:journals/dagstuhl-reports/AnandCJSS19}.
In conversational QA, a system is required to return a correct answer given a question and the previous conversation turns. Such questions are often ambiguous outside of the conversational context and require incorporating additional information contained in the previous conversation turns. Moreover, evaluating systems for conversational QA, especially for questions that require long generative answers drawn from multiple information sources, remains an open research problem in its own right~\cite{DBLP:conf/naacl/Voorhees03,DBLP:conf/naacl/KrishnaRI21,DBLP:conf/acl/SibliniSK20,li2021ditch}.

This paper provides an overview of the SCAI-QReCC 2021 shared task on conversational question answering. It reports on the extended conversational QA dataset employed in this task, participating systems and insights gained from their performance, and the challenges we faced when evaluating the submissions and our approach for dealing with those. Specifically, we designed and tested a set of guidelines to support evaluation of (conversational) QA models.

The shared task was built around the recently introduced QReCC dataset~\cite{anantha:2021}.
QReCC contains sequences of conversational questions paired with answers produced by human annotators.
Such sequences imitate a dialogue session with follow-up questions asked by the user.
The annotators had access to a web search engine and every answer is based on the content of a single web page, while several web pages may be used to answer different questions within the same sequence.
These web pages were downloaded and chunked into passages.
The task of the system is to retrieve information from this passage collection and produce the correct answer based on this information.

Since QReCC contains only one answer provided by the human annotators, our goal was not only to evaluate the current state-of-the-art but also to collect alternative correct answers for this dataset.
While the ground truth provides a single correct answer per question, in practice more than one answer can be considered correct.

\begin{figure}
\begin{center}
\includegraphics[scale=0.7]{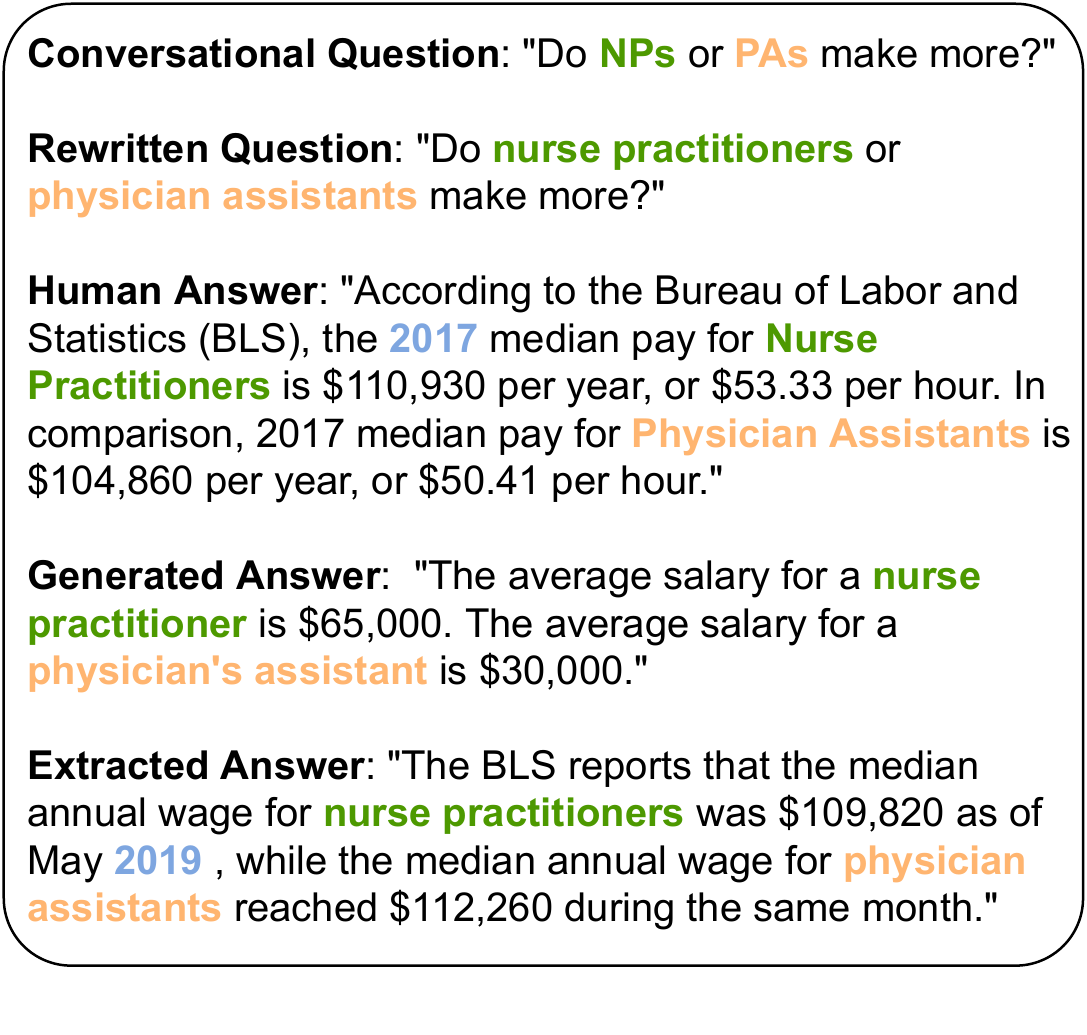}
\caption{A sample from the SCAI-QReCC results with the original conversational question, disambiguated rewritten question, a ground-truth human answer, an unfaithful generated answer from one of the participating systems and an alternative correct answer extracted from the submitted passages using our approach.
}
\label{fig:example}
\end{center}
\end{figure}

We received 29 runs, including the submissions made by four participating teams and the results produced by our three baseline models.
Each of these runs contains answers to all questions of the QReCC test set.
To assess these runs, we employed a range of automated metrics and arranged two crowdsourcing tasks.

Human assessment remains crucial for QA evaluation since automated measures may only assess similarity to the ground truth but the answer may be correct even when it differs from the ground truth.
Our goal was to find such answers and add them to the QReCC dataset.
Since it was not possible to manually evaluate all 17K  answers that we collected, we also had to come up with a set of techniques that allowed us to elicit a smaller subset that was more likely to contain alternative correct answers.

Another major challenge we faced was in judging answer correctness.
All teams participating in the shared task used generative models to produce the answers by conditioning on several previously retrieved passages.
However, it was previously shown that such models tend to deviate from the information provided in the passages~\cite{DBLP:conf/naacl/KrishnaRI21}.
While these generated answers may seem plausible to a human annotator, they may contain factual errors.

Considering the challenges mentioned above, we propose to evaluate the submitted answers in two stages: (1) answer plausibility: does it answer the question?; (2) answer faithfulness: does it follow from the evidence available from the passage collection?

Since the length of a single passage makes them unscrutinizable for a human evaluator, we employ a token-overlap-based heuristic to extract shorter spans from the retrieved passages.
Consider the example provided in Figure~\ref{fig:example}.
While the human answer is outdated and the generated answer is not factually correct, they both allow us to find similar answer spans within the retrieved passages that can be either used as evidence or as the extracted answers themselves.


The rest of the paper is structured as follows.
Section~\ref{experimental_setup} describes the subtasks, baselines and participating systems.
Section~\ref{automatic_evaluation} lists the metrics we used for automated evaluation and summarises their results.
In Section~\ref{human_evaluation} we provide a brief overview of our human evaluation procedure, which is then described in more detail in Sections~\ref{answer_plausibility}-\ref{answer_grounding}
Finally, Section~\ref{conclusion} concludes with a summary of our main findings, lessons learned and directions for future work.

\section{Task Setup}
\label{experimental_setup}


This section describes the setup of the 2021~edition of the SCAI-QReCC2021 shared task in more detail, including the three subtasks relevant for conversational QA, our baselines, and participating systems.

\subsection{Subtasks}
\label{experimental_setup_tasks}
Following the same setup as in QReCC, we decomposed the end-to-end conversational QA task into three subtasks that can be implemented and evaluated separately:
\begin{enumerate*}[label=(\arabic*)]
\item question rewriting -- ability of the model to correctly interpret and reformulate the question into its unambiguous equivalent;
\item passage retrieval -- ability of the model to locate information relevant for answering the question; and
\item answer generation -- ability of the model to produce faithful and grammatically correct answers.
\end{enumerate*}

While the conversational QA task can be approached end-to-end using a single model, such decomposition is beneficial for several reasons:
\begin{enumerate*}[label=(\arabic*)]
\item it allows to reuse the same passage retrieval and answer generation components already tuned for non-conversational QA~\cite{DBLP:conf/wsdm/VakulenkoLTA21};
\item it allows to efficiently scale retrieval to large document collections using sparse representation approaches~\cite{DBLP:journals/tacl/LuanETC21};
\item it enables a more thorough evaluation providing insights into the bottlenecks and opportunities to improve the end-to-end performance~\cite{vakulenko2020wrong}.
\end{enumerate*}



\subsection{Baselines}
\label{experimental_setup_baselines}
We introduced three baseline models to serve as reference points:

\begin{itemize}
\item \textit{Basic}.
This baseline implements a naive approach for question answering: it submits the question as the answer. Though not intuitive at first glance, this baseline is surprisingly hard to beat as most QA metrics are based on token overlap between submitted and ground truth answer, and the ground truth answers naturally often share tokens with the respective questions.

\item \textit{Simple}.
This baseline implements low-effort approaches for each subtask. For question rewriting, it submits the question without modification. For passage retrieval, it employs Pyserini%
\footnote{\footnoteurl{https://github.com/castorini/pyserini}}
BM25 with $k1=0.82$ and $b=0.68$ as in the QReCC paper~\cite{anantha:2021}. For question answering, it submits the sentence from the retrieved passages that contains the most of the stemmed noun phrases of the question. The baseline's source code is available on Github.%
\footnote{\footnoteurl{https://github.com/scai-conf/SCAI-QReCC-21/tree/main/code/simple-baseline}}

\item \textit{GPT3}.
This baseline uses the OpenAI GPT3 API~\cite{brown:2020} with default parameters (see Appendix) for question answering. The answers for 16,451 conversational questions of the test set were generated in 90~minutes for 33~USD. As the model prompt, all preceding questions and answers were prefixed with ``Q:'' and ``A:'' respectively and then concatenated.
\end{itemize}

\subsection{Participating Systems}
\label{experimental_setup_submissions}

We encouraged participants to submit working software through the TIRA~\cite{potthast:2019} platform but also allowed for traditional run file submissions. Each participant received access to a dedicated virtual machine with full admin rights and access to the QReCC dataset and a pre-build Anserini~\cite{yang:2017} index to deploy their software submissions and improve upon the available Pyserini baseline. We deployed the Pyserini baseline as software submission in TIRA and made the code available to simplify adaptations and encourage reproducibility because software submissions in TIRA can be executed on new datasets in the future without adoption. Still, all participants submitted run files and no working software because the deadline of the SCAI-QReCC~2021 shared task was close to the deadline of the TREC 2021~CAsT track.

Overall, four teams submitted results to the shared task:

\begin{itemize}
\item \textit{Rachael}~\cite{raposo:2022} implemented a three-stage pipeline that rewrites the question with T5~\cite{raffel:2019} and summarizes the top-10 passages retrieved with BM25 for the rewritten question using PEGASUS~\cite{zhang:2020b} to generate the answer. The T5 model for query rewriting is pre-trained\footnote{\footnoteurl{https://huggingface.co/castorini/t5-base-canard}} on the CANARD dataset~\cite{elgohary:2019} and uses the questions and answers of previous turns and the current question to make the current question context-independent. The rewritten query is used to retrieve the top-10 passages with BM25 implemented in Pyserini. Finally, the rewritten query and the top-10 passages are concatenated and fed to the abstractive summarizer PEGASUS fine-tuned on the official QReCC training set. The resulting summary is returned as the answer.

\item \textit{Rali-QA} used the human rewritten questions to retrieve passages with a pipeline identical to our simple baseline (BM25 with $k1=0.82$ and $b=0.68$) and a fine-tuned extractive BERT model for question answering on the top-k passages. The BERT model was fine-tuned for span extraction during a single epoch on the official QReCC training dataset, starting with the weights from a BERT model pre-trained on SQuAD v1.

\item \textit{Torch} uses a three-stage pipeline that rewrites the question with GPT2~\cite{radford:2019}, retrieves passages with BM25 and a BERT-based re-ranker, and generates the answer from the top-scored passage using T5~\cite{raffel:2019}. The question rewriting follows the idea of Yu et al.~\cite{yu:2020} and uses 
GPT2 fine-tuned on the official QReCC training set with the questions and answers of previous turns and the current question to rewrite the current question. The passage retrieval re-ranks the top-1000 results for the rewritten query of BM25 implemented in Anserini~\cite{yang:2017} ($k1=0.9$ and $b=0.4$) using the OpenMatch~\cite{liu2021b} BERT re-ranker pre-trained on MS-MARCO.\footnote{\footnoteurl{https://huggingface.co/castorini/monobert-large-msmarco}} Initially, the answer generation was intended in two stages using two T5~models. The first stage was intended to use a T5 model to generate a dedicated answer for each of the top-10 passages from the BERT re-ranking by concatenating the passage to the rewritten question. The second stage was intended to use another T5 model that uses the ten generated answers as input to generate the final answer as output. Due to the length of the passages, the fine-tuning of the second T5 model failed. Hence, the final answer was generated using the top-passage concatenated to the rewritten question (leaving the second stage of answer generation for future work).

\item \textit{Ultron} use a two-stage pipeline rewriting the question with the sequence-to-sequence model BART~\cite{lewis:2020} and generating the answers using RAG~\cite{lewis:2020a}. The BART model uses the current question and the queries of the previous turns as input to rewrite the current question. The BART model was fine-tuned for question rewriting using the official QReCC training set. Due to the large size of the collection, team ultron tested three different indexes for answer generation with RAG: (1) the Wikipedia index, (2) a filtered version of the QReCC passages using the top-10 passages for the questions of all turns of the conversation retrieved with BM25, and (3) a filtered version of the QReCC passages using the top-100 passages for the questions of all turns of the conversation retrieved with BM25.

\end{itemize}

\section{Automatic Evaluation}
\label{automatic_evaluation}

This section lists the metrics we used for automated evaluation and summarises their results.

\subsection{Metrics}
\label{automatic_evaluation_metrics}
The following section describes the metrics employed for each of the three subtasks. All metrics are averaged over all turns in the test set. Our script to calculate these metrics is made available as open-source.%
\footnote{\footnoteurl{https://github.com/scai-conf/SCAI-QReCC-21/tree/main/code/evaluation-script}. Due to the large model size, KPQA was computed using the original script in \footnoteurl{https://github.com/hwanheelee1993/KPQA}}

\paragraph{Question Rewriting}
We employ the measure which achieved in the dataset paper the highest correlation with human judgements for this subtask, ROUGE-1 \cite{anantha:2021}. The question rewriting subtask is only performed and evaluated on the {\it original} dataset as the questions in the {\it rewritten} dataset are already the ground-truth rewrites.
\begin{itemize}
\item {\it QR}. We report, ROUGE-1 \cite{lin:2004} here, which is the unigram recall (on token level) between the automatically rewritten questions and the ground-truth one as recommended based on the experimental evaluation performed by \cite{anantha:2021} (Section 6: Question Rewriting Metrics Validation).
\end{itemize}

\paragraph{Passage Retrieval}
From the dataset paper we adopt the use of the mean reciprocal rank to evaluate the ranking of passages retrieved for each question \cite{anantha:2021}. Although the use of MRR~faces some criticism~\cite{fuhr:2017,zobel:2020}, it is still used for evaluations in the TREC~2021 Conversational Assistance and Deep Learning tracks. We employ it here for comparability with previous work.
\begin{itemize}
\item {\it MRR}.
This metric is equal to $1 / r$, where $r$ is the rank of the highest-ranked relevant passage. We employ the token overlap heuristic of \newcite{anantha:2021} to determine relevance. Using token overlap is similar in spirit to several of the question answering metrics described below.
\end{itemize}

\paragraph{Question Answering}
We experiment with eight different metrics for comparing the generated answers with the ground-truth answers:
\begin{itemize}
\item {\it EM}.
The ``Exact Match'' is~1 if the answer is identical to the ground-truth after lowercasing, stemming, punctuation, and stopword removal. An Exact Match of~0 corresponds to disjoint pairs of text.
\item {\it F1}.
The F-Measure is the harmonic mean of precision and recall. These correspond here to the fraction of shared tokens between predicted and ground-truth answer among the tokens in the predicted or in the ground-truth answer, respectively.
\item {\it R1}.
ROUGE-1: the same as {\it QR} for question rewriting.
\item {\it POSS}.
The POSSCORE \cite{liu:2021} computes the cosine similarity of averaged word embeddings between predicted and ground-truth answers, but does so separately and weighted for tokens with specific part-of-speech tags and tokens with other tags. We use the default tag set: ADJ, ADV, VERB, PROPN, NOUN.
\item {\it SAS}.
Semantic Answer Similarity~\cite{risch:2021} uses a cross-encoder neural network, where both predicted and ground-truth answer are first concatenated with a separator token in between and then fed into a language model for similarity prediction.
\item {\it BERT}.
BERTScore \cite{zhang:2020} computes the token-wise F-Measure (see above) between predicted and ground-truth answers, but matches tokens based on the highest cosine similarity of the respective contextual BERT embeddings and uses this similarity instead of a binary exact match. Moreover, matches are weighed by the inverse document frequency of the matched token.
\item {\it BKPQA}/{\it RKPQA}.
BERTScore-KPQA and ROUGE-L-KPQA \cite{lee:2021} are modified version of BERTScore (see there) and ROUGE-L \cite{lin:2004} that weight each token by a predicted importance of the answer token with respect to the question. This importance score is predicted using a fully connected neural network that is trained on extractive question answering datasets.
\end{itemize}

\subsection{Results}
\label{automatic_evaluation_results}
The upper part of Table~\ref{table-scai-qrecc-21-scores} compares our baselines (Section~\ref{experimental_setup_baselines}) and the participating systems (Section~\ref{experimental_setup_submissions}) on the test split of the QReCC dataset.
The bottom part shows the scores when directly using the human rewritten questions as input instead of the original ones.

\begin{table*}
\centering\small\setlength{\tabcolsep}{1pt}
\begin{tabular}{@{}l@{\hspace{4pt}}lCCCCCCCCCCC@{}}
\toprule
\bf Team & \bf Run             & \bf QR    & \bf MRR   & \bf EM    & \bf F1    & \bf R1    & \bf POSS  & \bf SAS   & \bf BERT  & \bf BKPQA & \bf RKPQA \\
\midrule                                                                                                                                        
\multicolumn{12}{@{}l@{}}{\it Original questions}\\
\addlinespace[2pt]
-        & Basic baseline      & -         & -         & 0.000     & 0.114     & 0.095     & 1.283     & 0.207     & 0.422     & 0.432      & 0.064      \\
-        & GPT3 baseline       & -         & -         & 0.001     & 0.149     & 0.148     & 1.305     & 0.264     & 0.448     & 0.467      & 0.134      \\
-        & Simple baseline     & 0.571     & 0.065     & 0.001     & 0.067     & 0.150     & 1.490     & 0.162     & 0.367     & 0.426      & 0.097      \\
\addlinespace[2pt]
rachael  & 2021-09-04-10-38-07 & -         & 0.056     & 0.002     & 0.138     & 0.193     & \bf 1.583 & 0.163     & 0.410     & 0.476      & 0.135      \\
rachael  & 2021-09-08-07-07-57 & 0.675     & 0.135     & 0.006     & 0.187     & \bf 0.226 & 1.570     & \bf 0.277 & \bf 0.452 & \bf 0.498  & 0.175      \\
rachael  & 2021-09-08-07-09-57 & 0.682     & 0.128     & 0.006     & 0.186     & 0.226     & 1.558     & 0.269     & 0.448     & 0.494      & 0.175      \\
rachael  & 2021-09-08-15-40-34 & 0.679     & 0.133     & 0.007     & 0.176     & 0.211     & 1.456     & 0.254     & 0.420     & 0.460      & 0.164      \\
rachael  & 2021-09-08-21-49-44 & 0.681     & 0.130     & 0.008     & 0.177     & 0.211     & 1.461     & 0.246     & 0.422     & 0.462      & 0.167      \\
rachael  & 2021-09-15-09-05-06 & 0.673     & \bf 0.158 & \bf 0.011 & 0.179     & 0.212     & 1.333     & 0.254     & 0.405     & 0.444      & 0.172      \\
rachael  & 2021-09-15-09-06-44 & 0.681     & 0.150     & 0.010     & 0.179     & 0.211     & 1.369     & 0.249     & 0.408     & 0.449      & 0.169      \\
rachael  & 2021-09-15-09-07-49 & 0.676     & 0.157     & 0.010     & 0.187     & 0.219     & 1.399     & 0.264     & 0.418     & 0.457      & 0.175      \\
rachael  & 2021-09-15-09-08-40 & \bf 0.685 & 0.149     & 0.010     & \bf 0.189 & 0.222     & 1.458     & 0.259     & 0.428     & 0.470      & \bf 0.178  \\
\addlinespace[2pt]
torch    & usi$\_$T5$\_$raw2   & 0.657     & 0.082     & 0.001     & 0.137     & 0.200     & 1.451     & 0.221     & 0.415     & 0.467      & 0.117      \\
\midrule
\multicolumn{12}{@{}l@{}}{\it Human rewritten questions}\\
\addlinespace[2pt]
-        & \multicolumn{2}{@{}l}{Basic baseline}                        & -         & 0.000     & 0.224     & 0.205     & 1.555     & 0.351     & 0.517     & 0.472      & 0.132      \\
-        & \multicolumn{2}{@{}l}{Simple baseline}                       & \bf 0.398 & 0.001     & 0.098     & 0.282     & 1.666     & 0.372     & -         & -          & -          \\
\addlinespace[2pt]
rachael  & \multicolumn{2}{@{}l}{2021-09-04-10-39-42}                   & 0.359     & 0.011     & 0.267     & 0.331     & \bf 1.674 & 0.398     & 0.534     & 0.562      & 0.258      \\
rachael  & \multicolumn{2}{@{}l}{2021-09-06-09-21-43}                   & 0.359     & 0.018     & 0.290     & 0.339     & 1.649     & \bf 0.430 & \bf 0.549 & \bf 0.570  & 0.277      \\
rachael  & \multicolumn{2}{@{}l}{2021-09-15-19-36-31}                   & 0.385     & \bf 0.028 & \bf 0.302 & \bf 0.345 & 1.618     & 0.420     & 0.544     & 0.566      & \bf 0.290  \\
\addlinespace[2pt]
rali-qa  & \multicolumn{2}{@{}l}{2021-09-09-13-01-07}                   & 0.269     & 0.003     & 0.166     & 0.212     & 1.385     & 0.264     & 0.407     & 0.457      & 0.174      \\
\addlinespace[2pt]
ultron   & \multicolumn{2}{@{}l}{2021-09-04-17-28-07}                   & -         & 0.001     & 0.183     & 0.186     & 1.357     & 0.301     & 0.463     & 0.457      & 0.121      \\
ultron   & \multicolumn{2}{@{}l}{2021-09-08-15-04-28}                   & -         & 0.015     & 0.261     & 0.258     & 1.565     & 0.383     & 0.533     & 0.539      & 0.220      \\
ultron   & \multicolumn{2}{@{}l}{2021-09-08-15-07-30}                   & -         & 0.001     & 0.187     & 0.189     & 1.380     & 0.306     & 0.472     & 0.465      & 0.123      \\
ultron   & \multicolumn{2}{@{}l}{2021-09-08-15-08-00}                   & -         & 0.004     & 0.247     & 0.236     & 1.597     & 0.379     & 0.536     & 0.525      & 0.177      \\
ultron   & \multicolumn{2}{@{}l}{bart-large$\_$top1bm25}                & -         & 0.000     & 0.017     & 0.017     & 0.150     & 0.111     & 0.046     & 0.048      & 0.016      \\
ultron   & \multicolumn{2}{@{}l}{distilbart-xsum-12-1$\_$top1bm25}      & -         & 0.000     & 0.019     & 0.020     & 0.170     & 0.113     & 0.050     & 0.054      & 0.018      \\
ultron   & \multicolumn{2}{@{}l}{distilbart-xsum-12-3$\_$top1bm25}      & -         & 0.000     & 0.022     & 0.023     & 0.175     & 0.117     & 0.052     & 0.056      & 0.021      \\
ultron   & \multicolumn{2}{@{}l}{rag-bm25$\_$100}                       & -         & 0.004     & 0.247     & 0.236     & 1.597     & 0.379     & 0.536     & 0.525      & 0.177      \\
ultron   & \multicolumn{2}{@{}l}{rag-dpr-hard-neg-bm25-top10}           & -         & 0.015     & 0.261     & 0.258     & 1.565     & 0.383     & 0.533     & 0.539      & 0.220      \\
ultron   & \multicolumn{2}{@{}l}{rag-ft-dpr-hard-neg-bm25$\_$10} & -         & 0.015     & 0.261     & 0.258     & 1.565     & 0.383     & 0.533     & 0.539      & 0.220      \\
\bottomrule
\end{tabular}
\caption{Evaluation results on the original dataset (top) and when using the human rewritten questions as input. Metrics are described in Section~\ref{automatic_evaluation_metrics}. A "-" denotes that no output was submitted for the respective task or the evaluation code failed (for BERT, BKPQA, and RKPQA).}
\label{table-scai-qrecc-21-scores}
\end{table*}

\paragraph{Question Rewriting}
This subtask is evaluated on the original dataset only. The QR column shows the ROUGE-1 that the approaches reached. The simple baseline returns the question as-is and is outperformed by every run of the two teams that implemented their own rewriting approaches. Both teams achieved similar scores, with team Rachael having a slight advantage indicating that T5 might be a good starting point for further improving question rewriting.

\begin{figure}
\includegraphics[width=\linewidth]{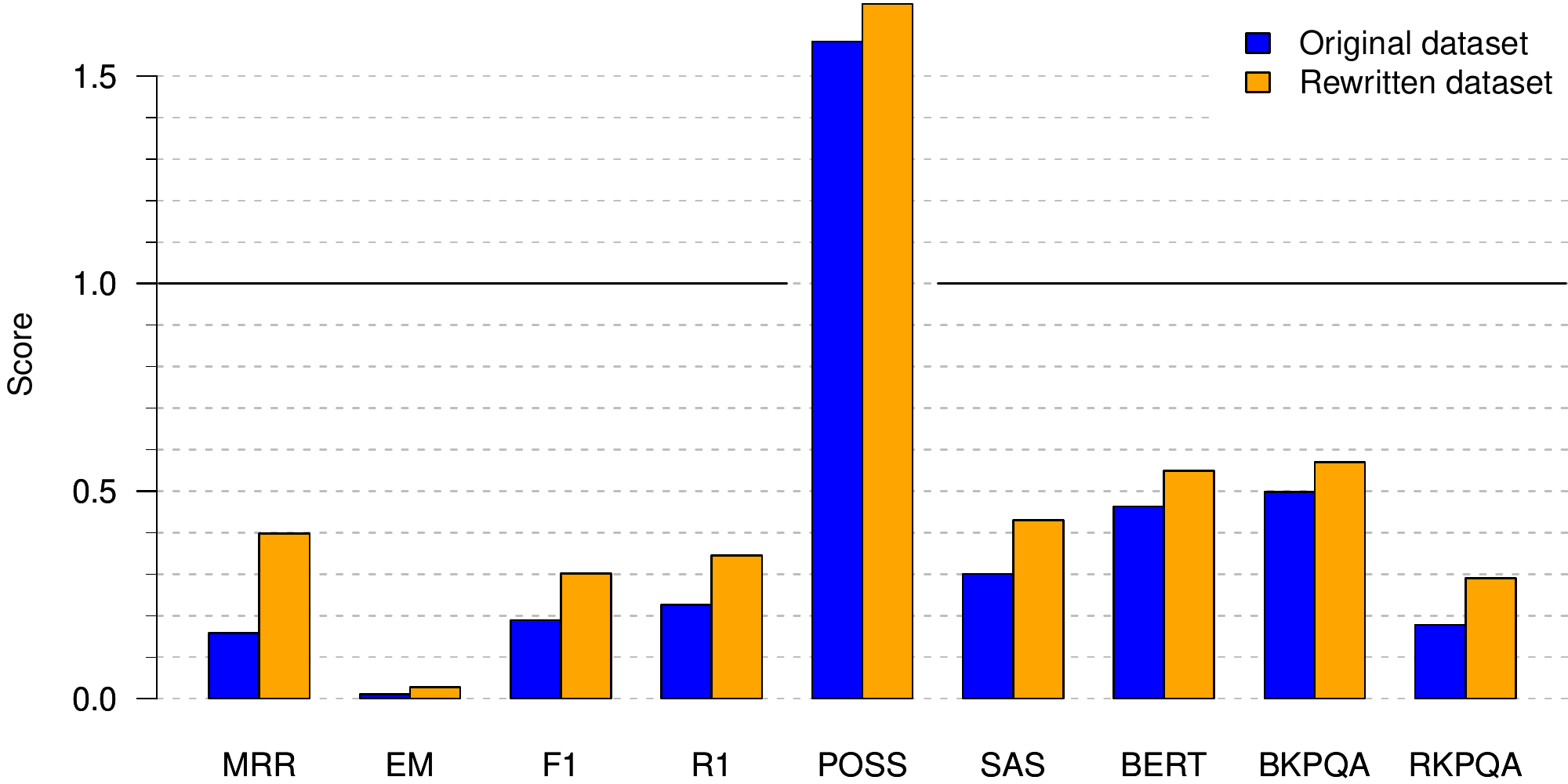}
\caption{Highest scores reached per dataset variant for each metric. For all metrics except POSS a score of 1 corresponds to optimal performance.}
\label{plot-best-run-by-metric}
\end{figure}

\paragraph{Passage Retrieval}
The teams Rachael, Rali-QA, and Torch submitted results for the passage retrieval subtask, which we compared in terms of the mean reciprocal rank (MRR). Nearly all submitted runs improve upon the simple baseline, which uses the BM25-retrieval model with a standard parameter set. The best run, by team Rachael, reaches even a more than twice as high score. However, the run with the highest QR does not reach the highest MRR. Indeed, the Pearson correlation of a runs QR and MRR for team Rachael is~-0.35, indicating a weak negative correlation. The scores for QR and MRR are similar across team Rachael's runs, though. In general, a considerably higher MRR is reached when using human rewritten questions, showing the necessity of the rewriting step. Figure~\ref{plot-best-run-by-metric} visualizes this performance increase. Surprisingly, the simple baseline performed best in this case. A possible explanation is that the not-rewritten questions it uses provide a much worse starting point for passage retrieval than the automatic rewritten questions of the participating teams.

\paragraph{Question Answering}
All four teams participated in the question answering subtask. For this task, the baselines are consistently beaten by at least one participant for each metric. However, we find that the rankings of the different metrics often disagree, leaving otherwise an inconclusive image. As an extreme case, the runs by team Rachael that achieved the highest POSSCORE (for original and human rewritten questions) actually got the lowest score of their runs for most other metrics. The only exceptions are the BERTScore and BERTScore-KPQA metric when using original questions. Despite these differences, however, the metrics generally agree on ranking the runs of team Rachael high up, indicating the potential of their approach. Moreover, as Figure~\ref{plot-best-run-by-metric} shows, using human rewritten questions instead of the original ones does also improve question answering performance---for all metrics---, though to a lesser extent than for passage retrieval (MRR).
To enrich these automated results and provide more insight, we also employed a human evaluation procedure outlined in the next section.

\section{Human Evaluation}
\label{human_evaluation}
\begin{figure}
\includegraphics[width=\linewidth]{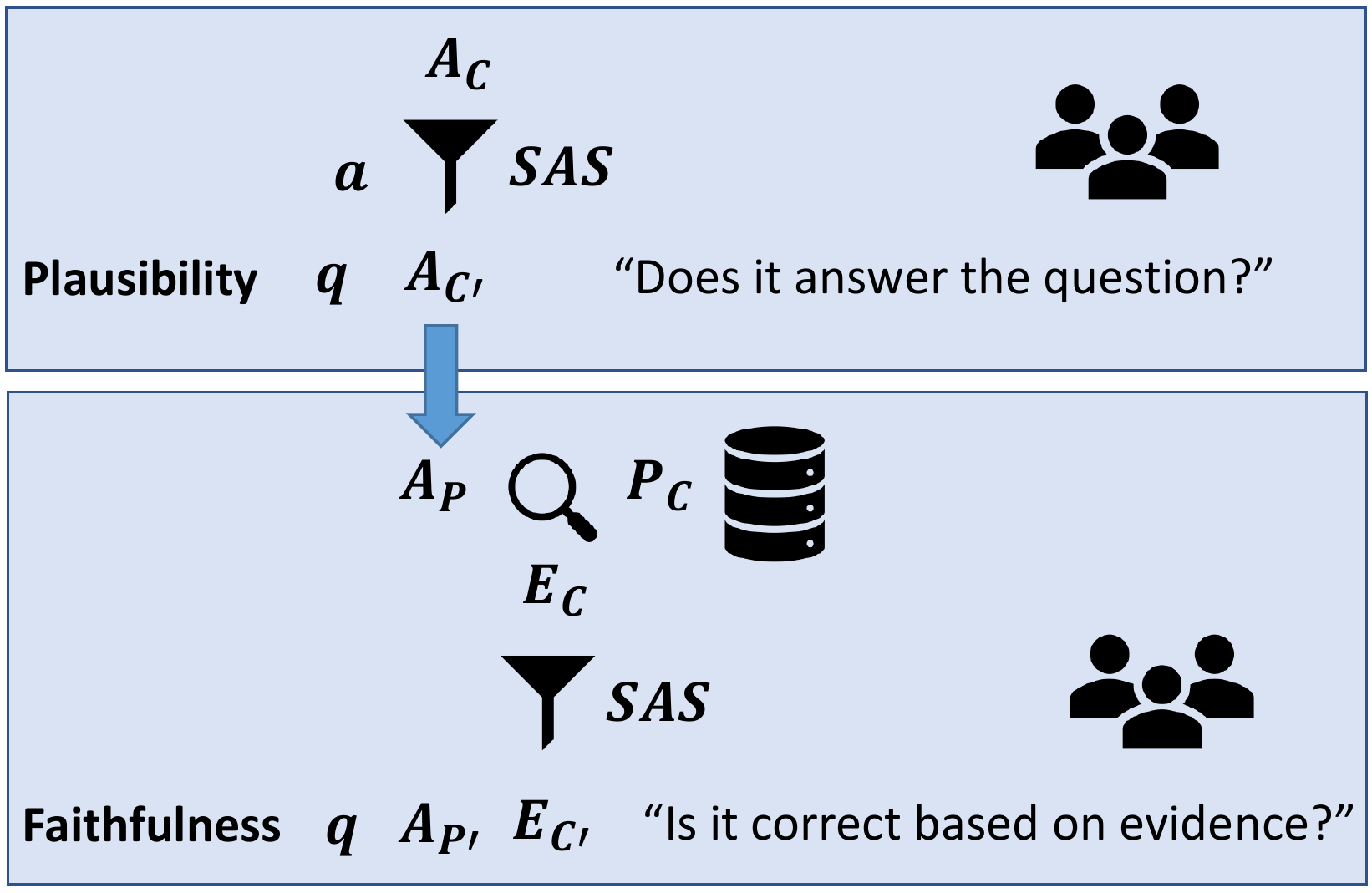}
\caption{Our human evaluation including assessment of the answer plausibility and answer faithfulness.}
\label{overview}
\end{figure}

An overview of our human evaluation of the QA performance is given in Figure~\ref{overview} and consists of two phases.
We start with the set of candidate answers $A_C$ for question $q$ that was obtained from the runs submitted by the participants and our baseline approaches.
Then, we use the SAS score~\cite{risch:2021} that shows similarity to the ground truth answer $a$ from QReCC to filter out the candidate answers that are more likely to be correct.
The subset with the highest SAS scores $A_{C'} \subseteq A_C$ is then used to crowdsource answer plausibility labels (see Section~\ref{answer_plausibility} for more details).
The result of this phase is the subset $A_{P} \subseteq A_{C'}$ of all plausible answers identified for question $q$.

The next evaluation phase is designed to assess faithfulness of the plausible answers identified at the previous phase.
For each plausible answer in $A_{P}$, we collect a set of evidence spans $E_C$ by matching each answer to the top-k passages submitted by the participants $P_C$ using a sliding window and a token-overlap heuristic.
We keep the token-overlap threshold low to find semantic matches and then apply the SAS score to filter out the candidate spans.
The resulting sets of spans extracted from the passages $E_{C'} \subseteq E_C$ paired with the corresponding plausible answers $A_{P'} \subseteq A_{P}$ is then used to crowdsource faithfulness labels (see Section~\ref{answer_grounding} for more details).

\section{Answer Plausibility}
\label{answer_plausibility}

In this section, we describe our answer sampling approach and annotation instructions.
We conclude with summarising our results. 

\subsection{Answer sampling}
We obtained 16,736 answers, in total, from all the submitted runs including our baselines and participant systems.
Our goal was to find alternative correct answers or answer variations that could help us to extend the ground-truth answers in QReCC.
We also decided to focus on the questions on which the participating systems tend to disagree.
To this end, we sampled at least four alternative answers per question submitted by the systems that are all distinct but at the same time semantically close to the ground truth answer using the SAS score (described in Section~\ref{automatic_evaluation_metrics}) with the threshold above 0.7.
While the SAS score allows to find answer paraphrases beyond the lexical overlap matches, it also enables us to capture alternative answers that may contradict the ground truth answers (see the Generated Answer in Figure~\ref{fig:example}).

\subsection{Annotation task}

\begin{figure}[!h]
\begin{center}
\fbox{\includegraphics[scale=0.27]{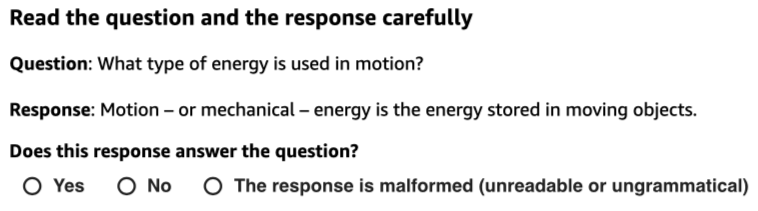}}
\caption{Task setup for answer plausibility annotation.}
\label{fig:answer_plausibility}
\end{center}
\end{figure}

We asked the crowd workers using Amazon Mechanical Turk (MTurk) to judge the answers with respect to the questions (see Figure~\ref{fig:answer_plausibility}).
This annotation task was estimated to take about 30 seconds for a single QA pair, on average.
The workers were reimbursed with \$0.06 per sample they annotated.\footnote{The rate was calculated based on an hourly wage of \$7.25, which is the US federal minimum wage provided by the US Department of Labor.}

We started the crowdsourcing experiment with several smaller batches at first and performed the quality assurance of the results to select a pool of the trusted crowd workers who then completed the rest of the annotations.
One of the authors did the quality assurance by manually examining the samples with disagreements.
We collected two annotations per sample.
The disagreements were manually resolved for the first batches.
The rest of samples on which our pool of trust crowd workers disagreed were automatically discarded.

\subsection{Results}
\begin{figure}
\begin{center}
\includegraphics[scale=0.7]{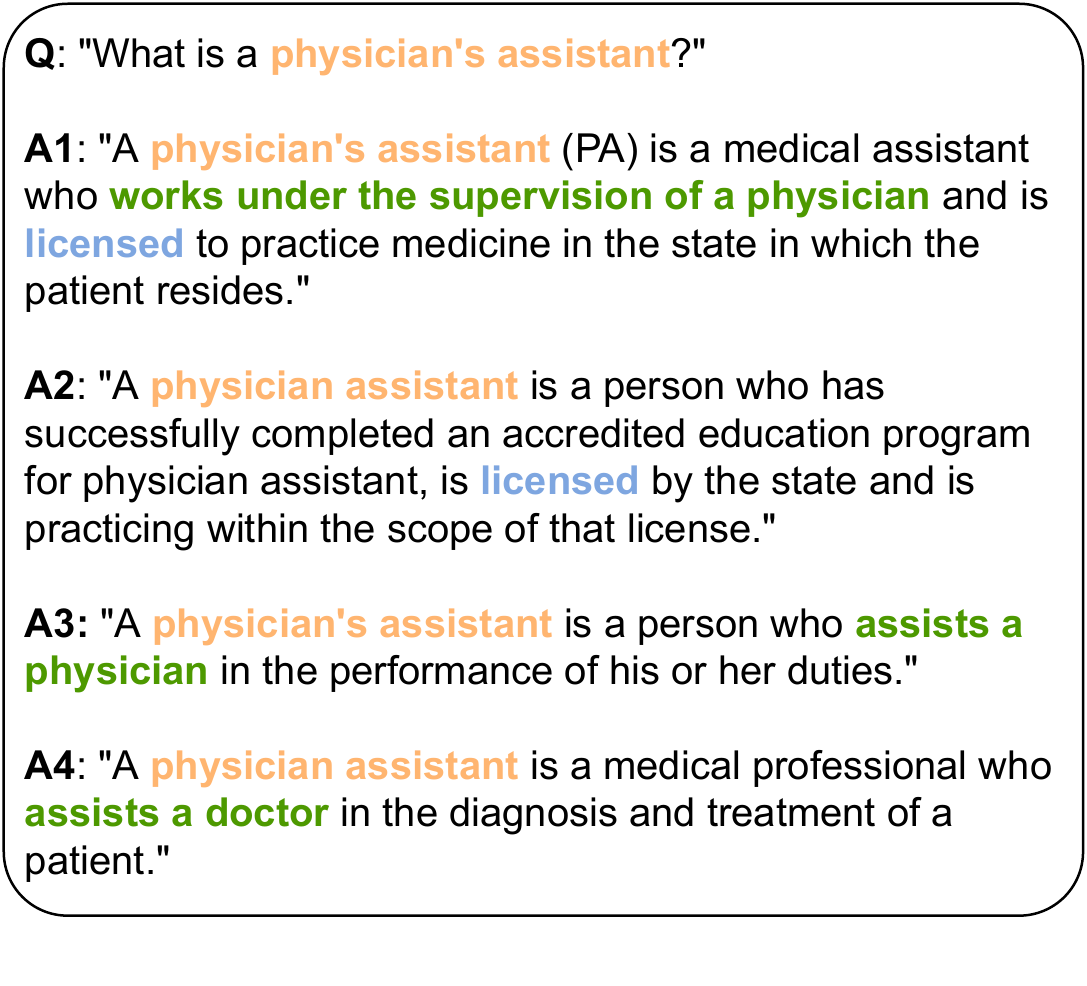}
\caption{A sample cluster of plausible answers for the same question from the SCAI-QReCC results.}
\label{fig:example2}
\end{center}
\end{figure}

After processing the annotations, we collected 1,863 answers judged as plausible, 108 malformed answers, and 107 implausible answers (see Table~\ref{table:human_eval} for per run distribution).
Thereby, we obtained 465 clusters with more than one plausible answers for the same question (avg: 4, max: 13 answers per cluster).
In some cases those answers were paraphrases, shortened or more detailed versions of the same answer.
In other cases, the answers are different, which may or may not be contradictory, such as pointing at different aspects in a definition (see the example provided in Figure~\ref{fig:example2}).

\section{Answer Faithfulness}
\label{answer_grounding}
This section describes our answer grounding approach that produces short evidence spans enabling the crowd-workers to efficiently judge answer faithfulness.

\subsection{Evidence sampling}
Evaluating answer correctness with respect to the passage collection (faithfulness) is a hard task because the passages are very long for crowd workers to go through.
Our approach to sampling evidence spans is based on the observation that the human answers often paraphrase the original paragraph text in QReCC but, in the majority of cases, those paraphrases stay sufficiently close to the original such that the token overlap heuristic helps to identify them.

Firstly, we use the same span heuristic proposed in \cite{anantha:2021} to select short text spans within the passages with the maximum token overlap.
We lower the token overlap threshold considerably to allow for non-verbal semantic matches as well.
The evidence spans were trimmed to contain full sentences for readability.
Secondly, we use the SAS score to compare between the matched sentences and all the generated answers to select a pool of evidence for each question.
Similarly, we also sample additional evidence using the ground-truth answers from QReCC produced by human annotators.

Afterwards, we just merge all the evidence spans detected this way into a single paragraph.
In this manner, we still attempt to evaluate the answers for cases when the collected spans may already contain sufficient evidence  to judge the answer faithfulness even though they did not match any of the evidence spans directly.

\subsection{Annotation task}
\begin{figure}[!h]
\begin{center}
\fbox{\includegraphics[scale=0.32]{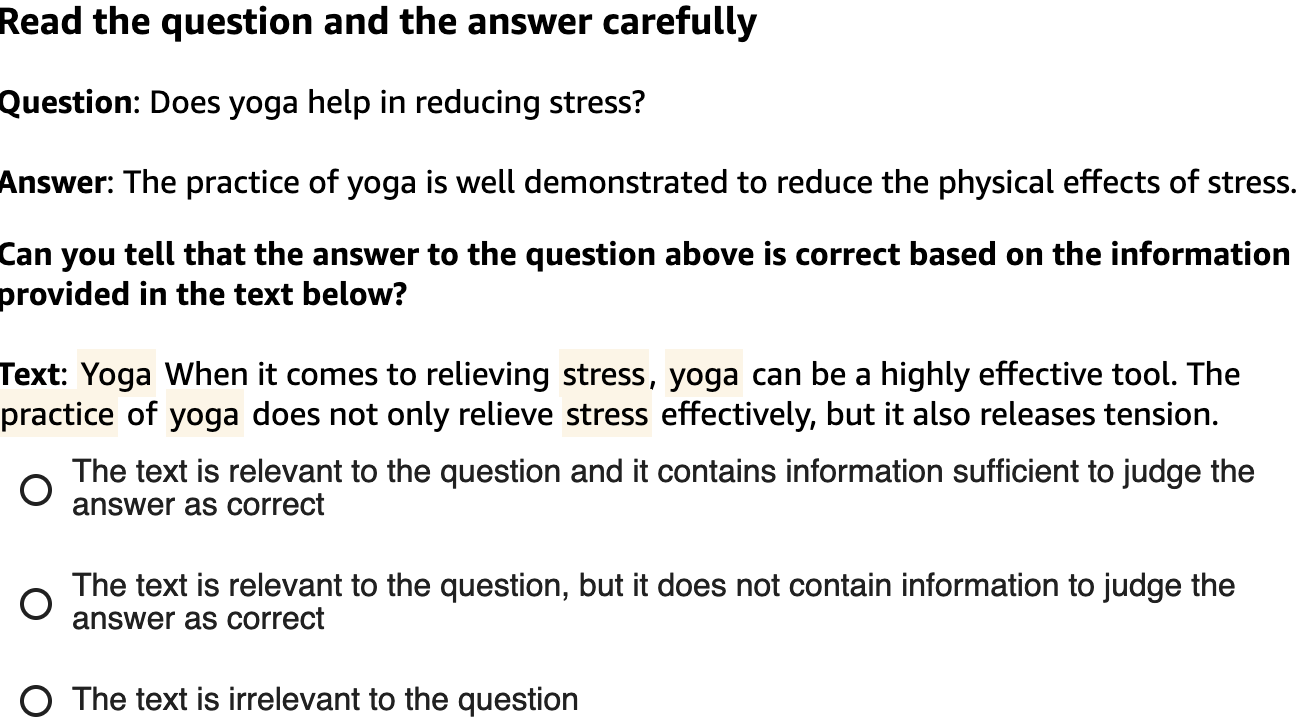}}
\caption{Task setup for answer grounding annotation with evidence sentences sampled from several retrieved passages by matching to the generated and ground-truth answers.}
\label{fig:answer_grounding}
\end{center}
\end{figure}

We annotated two batches with 578 samples, in total.
Each sample contains a triple of a question, one of the plausible answers and the text, which is a concatenation of all the evidence spans obtained for this question (see Figure~\ref{fig:answer_grounding} for an example of one such sample).
The workers were reimbursed \$0.12 per sample since they had to read through the answers and the matched evidence spans as well.
We followed the setup similar to the previous annotation task with two workers annotating the same sample.
We also used the same pool of MTurk crowd workers that was selected during the previous annotation task.

\subsection{Results}
After removing the samples on which the workers disagreed, we obtained 386 answers that were judged as faithful given the matched evidence spans.
Our results are summarised in Table~\ref{table:human_eval}.
Most of the answers were judged as faithful given the evidence spans we extracted, which shows the effectiveness of our evidence extraction approach and fidelity of the evaluated models.
The table also demonstrates that some of the models that produced many plausible answers, such as GPT3, have a lower proportion of answers judged as faithful than other models.
Note that this may also indicate that the answers generated by these models are very different from the retrieved passages.
In this case, our approach is not sufficient to detect relevant evidence spans.
Therefore, we believe that a reasonable requirement for generative QA models in the future shared tasks should be to provide the short evidence spans alongside with the passage IDs that can be used for answer evaluation.


\begin{table*}
\centering\small\setlength{\tabcolsep}{4pt}
\begin{tabular}{@{}lll@{\hspace{8\tabcolsep}}ccccc@{}}
\toprule
\bf Team & \bf Run & \bf Question & \bf Plausible & \bf Implausible & \bf Malformed & \bf Faithful & \bf Unfaithful \\
\midrule
rachael & 2021-09-04-10-39-42 & rewritten & \phantom{0}\textbf{183} & \phantom{00}5 & \phantom{00}4 & \phantom{0}\textbf{37} & \phantom{0}2 \\
rachael & 2021-09-08-21-49-44 & original  & \phantom{0}133 & \phantom{00}6 & \phantom{00}4 & \phantom{0}30 & \phantom{0}1 \\
rachael & 2021-09-08-07-07-57 & original  & \phantom{0}120 & \phantom{00}4 & \phantom{00}5 & \phantom{0}30 & \phantom{0}0 \\
rachael & 2021-09-15-09-07-49 & original  & \phantom{0}103 & \phantom{00}4 & \phantom{00}6 & \phantom{0}29 & \phantom{0}1 \\
-       & GPT3 baseline       & original  & \phantom{0}149 & \phantom{00}4 & \phantom{00}8 & \phantom{0}28 & \phantom{0}3 \\
ultron  & rag-bm25\_100       & rewritten & \phantom{0}173 & \phantom{0}15 & \phantom{00}6 & \phantom{0}27 & \phantom{0}2 \\
rachael & 2021-09-06-09-21-43 & rewritten & \phantom{0}158 & \phantom{00}4 & \phantom{00}3 & \phantom{0}26 & \phantom{0}\textbf{4} \\
ultron  & 2021-09-08-15-04-28 & rewritten & \phantom{0}149 & \phantom{0}\textbf{16} & \phantom{00}6 & \phantom{0}24 & \phantom{0}1 \\
rachael & 2021-09-15-19-36-31 & rewritten & \phantom{0}132 & \phantom{00}2 & \phantom{00}2 & \phantom{0}24 & \phantom{0}0 \\
rachael & 2021-09-15-09-06-44 & original  & \phantom{00}73 & \phantom{00}0 & \phantom{00}4 & \phantom{0}22 & \phantom{0}1 \\
rachael & 2021-09-08-07-09-57 & original  & \phantom{00}75 & \phantom{00}2 & \phantom{00}4 & \phantom{0}16 & \phantom{0}1 \\
rali-qa & 2021-09-09-13-01-07 & rewritten & \phantom{00}33 & \phantom{00}6 & \phantom{0}11 & \phantom{0}16 & \phantom{0}1 \\
rachael & 2021-09-08-15-40-34 & original  & \phantom{00}41 & \phantom{00}6 & \phantom{00}2 & \phantom{0}14 & \phantom{0}3 \\
torch   & usi T5 raw2         & original  & \phantom{00}36 & \phantom{00}7 & \phantom{0}\textbf{16} & \phantom{0}14 & \phantom{0}0 \\
ultron  & 2021-09-04-17-28-07 & rewritten & \phantom{0}117 & \phantom{0}13 & \phantom{00}7 & \phantom{0}13 & \phantom{0}0 \\
rachael & 2021-09-15-09-08-40 & original  & \phantom{00}52 & \phantom{00}4 & \phantom{00}4 & \phantom{0}10 & \phantom{0}0 \\
ultron  & BART-large-top1BM25 & rewritten & \phantom{00}29 & \phantom{00}3 & \phantom{0}11 & \phantom{0}10 & \phantom{0}0 \\
rachael & 2021-09-15-09-05-06 & original  & \phantom{00}52 & \phantom{00}2 & \phantom{00}1 & \phantom{00}9 & \phantom{0}0 \\
rachael & 2021-09-04-10-38-07 & original  & \phantom{00}41 & \phantom{00}2 & \phantom{00}0 & \phantom{00}6 & \phantom{0}1 \\
-       & Simple baseline     & rewritten & \phantom{00}14 & \phantom{00}2 & \phantom{00}3 & \phantom{00}1 & \phantom{0}0 \\
-       & Simple baseline     & original  & \phantom{000}0 & \phantom{00}0 & \phantom{00}1 & \phantom{00}0 & \phantom{0}0 \\
\midrule
Total   &                     &           & 1863 & 107 & 108 & 386 & 21 \\
\bottomrule
\end{tabular}
\caption{Human evaluation results of answer plausibility and faithfulness. The runs are ordered by the number of faithful answers. The highest values in each of the columns are highlighted in \textbf{bold}.}
\label{table:human_eval}
\end{table*}

\section{Conclusion}
\label{conclusion}
Results of the SCAI-QReCC shared task identified main achievements as well as the major challenges when applying the state-of-the-art models to the task of open-doman QA.
All of the submitted runs used a sparse index with BM25 for initial retrieval in combination with question rewriting for conversational QA.
Due to the large collection size, none of the participant teams managed to scale dense passage retrieval that could allow to deploy an end-to-end conversational QA model.
These results provide an important indicator of the technology maturity level for large-scale QA and conversational QA beyond Wikipedia-sized corpora. 


Overall, we proposed and tested in practice an evaluation procedure that allowed us to compare the model performance and discover new plausible and faithful answers.
We used it to extend the original conversational QA dataset used for evaluation with multiple correct answers per question.
While it is impossible to elicit a complete list of all correct answers for a given question, especially for an open-ended non-factoid question, this dataset is designed to improve our understanding of answer variations and specific properties important for a good answer.
Our dataset is made available under the public URL: \url{https://doi.org/10.5281/zenodo.5749472}

Our evaluation results showed that the modern QA models are already able to produce fluent answers but we can not always trust those answers to be correct.
Ordinary users are unaware of such models' limitations and can be easily persuaded or misguided by plausible but unfaithful answers.
There’s a need to establish a way to produce high quality answers grounded in the external information sources that can be referenced by the QA model. 
While we showed an efficient approach to mine plausible and faithful answers, it is not possible to evaluate faithfulness if the generated answers are very different from the original text.
To make further progress on the generative QA task, the models should be required to provide evidence alongside with their answers explicitly indicating the answer provenance.
Such evidence should be also limited to relatively short spans of bounded length, such that they can be easily examined and assessed by human evaluators.







\section{Bibliographical References}
\bibliographystyle{lrec2022-bib}
\bibliography{lrec22-scai-qrecc-overview-lit}

\begin{thebibliography}{}

\bibitem[\protect\citename{{McEnery A. et. al.}}2004]{EMILLE}
{McEnery A. et. al.}
\newblock (2004).
\newblock {\em The EMILLE/CIIL Corpus}.
\newblock distributed via ELRA: ELRA-Id W0037, ISLRN 039-846-040-604-0.

\bibitem[\protect\citename{{Speecon Consortium}}2011]{Speecon}
{Speecon Consortium}.
\newblock (2011).
\newblock {\em Catalan Speecon database}.
\newblock Speecon Project, distributed via ELRA: ELRA-Id S0327, Speecon
  resources, 1.0, ISLRN 935-211-147-357-5.

\end{thebibliography}


\begin{thebibliography}{}

\bibitem[\protect\citename{Castor and Pollux}1992]{CastorPollux-92}
Castor, A. and Pollux, L.~E.
\newblock (1992).
\newblock The use of user modelling to guide inference and learning.
\newblock {\em Applied Intelligence}, 2(1):37--53.

\bibitem[\protect\citename{Strötgen and Gertz}2012]{Martin-90}
Strötgen, J. and Gertz, M.
\newblock (2012).
\newblock Temporal tagging on different domains: Challenges, strategies, and
  gold standards.
\newblock In Nicoletta Calzolari~(Conference Chair), et~al., editors, {\em
  Proceedings of the Eight International Conference on Language Resources and
  Evaluation (LREC'12)}, pages 3746--3753, Istanbul, Turkey, may. European
  Language Resource Association (ELRA).

\bibitem[\protect\citename{Superman \bgroup et al.\egroup
  }2000]{Superman-Batman-Catwoman-Spiderman-00}
Superman, S., Batman, B., Catwoman, C., and Spiderman, S.
\newblock (2000).
\newblock {\em Superheroes experiences with books}.
\newblock The Phantom Editors Associates, Gotham City, 20th edition.

\end{thebibliography}


\begin{thebibliography}{}

\bibitem[\protect\citename{Anand \bgroup et al.\egroup
  }2019]{DBLP:journals/dagstuhl-reports/AnandCJSS19}
Anand, A., Cavedon, L., Joho, H., Sanderson, M., and Stein, B.
\newblock (2019).
\newblock Conversational search (dagstuhl seminar 19461).
\newblock {\em Dagstuhl Reports}, 9(11):34--83.

\bibitem[\protect\citename{Anantha \bgroup et al.\egroup }2021]{anantha:2021}
Anantha, R., Vakulenko, S., Tu, Z., Longpre, S., Pulman, S., and Chappidi, S.
\newblock (2021).
\newblock Open-domain question answering goes conversational via question
  rewriting.
\newblock In Kristina Toutanova, et~al., editors, {\em Conference of the North
  American Chapter of the Association for Computational Linguistics: Human
  Language Technologies (NAACL-HLT'21)}, pages 520--534. Association for
  Computational Linguistics.

\bibitem[\protect\citename{Brown \bgroup et al.\egroup }2020]{brown:2020}
Brown, T.~B., Mann, B., Ryder, N., Subbiah, M., Kaplan, J., Dhariwal, P.,
  Neelakantan, A., Shyam, P., Sastry, G., Askell, A., Agarwal, S.,
  Herbert{-}Voss, A., Krueger, G., Henighan, T., Child, R., Ramesh, A.,
  Ziegler, D.~M., Wu, J., Winter, C., Hesse, C., Chen, M., Sigler, E., Litwin,
  M., Gray, S., Chess, B., Clark, J., Berner, C., McCandlish, S., Radford, A.,
  Sutskever, I., and Amodei, D.
\newblock (2020).
\newblock Language models are few-shot learners.
\newblock In Hugo Larochelle, et~al., editors, {\em 33rd Annual Conference on
  Neural Information Processing Systems (NeurIPS'20)}.

\bibitem[\protect\citename{Elgohary \bgroup et al.\egroup }2019]{elgohary:2019}
Elgohary, A., Peskov, D., and Boyd{-}Graber, J.~L.
\newblock (2019).
\newblock Can you unpack that? learning to rewrite questions-in-context.
\newblock In Kentaro Inui, et~al., editors, {\em Proceedings of the 2019
  Conference on Empirical Methods in Natural Language Processing and the 9th
  International Joint Conference on Natural Language Processing, {EMNLP-IJCNLP}
  2019, Hong Kong, China, November 3-7, 2019}, pages 5917--5923. Association
  for Computational Linguistics.

\bibitem[\protect\citename{Fuhr}2017]{fuhr:2017}
Fuhr, N.
\newblock (2017).
\newblock Some common mistakes in {IR} evaluation, and how they can be avoided.
\newblock {\em {SIGIR} Forum}, 51(3):32--41.

\bibitem[\protect\citename{Gonçalo~Raposo and Coheur}2022]{raposo:2022}
Gonçalo~Raposo, Rui~Ribeiro, B.~M. and Coheur, L.
\newblock (2022).
\newblock {Question rewriting? Assessing its importance for conversational
  question answering}.
\newblock In {\em Advances in Information Retrieval. 44th European Conference
  on IR Research (ECIR 2022)}, Lecture Notes in Computer Science, Berlin
  Heidelberg New York, March. Springer.

\bibitem[\protect\citename{Kim \bgroup et al.\egroup
  }2021]{DBLP:conf/acl/KimKPK20}
Kim, G., Kim, H., Park, J., and Kang, J.
\newblock (2021).
\newblock Learn to resolve conversational dependency: {A} consistency training
  framework for conversational question answering.
\newblock In {\em Proceedings of the 59th Annual Meeting of the Association for
  Computational Linguistics and the 11th International Joint Conference on
  Natural Language Processing, {ACL/IJCNLP} 2021, (Volume 1: Long Papers),
  Virtual Event, August 1-6, 2021}, pages 6130--6141. Association for
  Computational Linguistics.

\bibitem[\protect\citename{Krishna \bgroup et al.\egroup
  }2021]{DBLP:conf/naacl/KrishnaRI21}
Krishna, K., Roy, A., and Iyyer, M.
\newblock (2021).
\newblock Hurdles to progress in long-form question answering.
\newblock In {\em Proceedings of the 2021 Conference of the North American
  Chapter of the Association for Computational Linguistics: Human Language
  Technologies, {NAACL-HLT} 2021, Online, June 6-11, 2021}, pages 4940--4957.
  Association for Computational Linguistics.

\bibitem[\protect\citename{Lee \bgroup et al.\egroup }2021]{lee:2021}
Lee, H., Yoon, S., Dernoncourt, F., Kim, D.~S., Bui, T., Shin, J., and Jung, K.
\newblock (2021).
\newblock {KPQA:} {A} metric for generative question answering using keyphrase
  weights.
\newblock In Kristina Toutanova, et~al., editors, {\em Conference of the North
  American Chapter of the Association for Computational Linguistics: Human
  Language Technologies (NAACL-HLT'21)}, pages 2105--2115. Association for
  Computational Linguistics.

\bibitem[\protect\citename{Lewis \bgroup et al.\egroup }2020a]{lewis:2020}
Lewis, M., Liu, Y., Goyal, N., Ghazvininejad, M., Mohamed, A., Levy, O.,
  Stoyanov, V., and Zettlemoyer, L.
\newblock (2020a).
\newblock {BART:} denoising sequence-to-sequence pre-training for natural
  language generation, translation, and comprehension.
\newblock In {\em Proceedings of the 58th Annual Meeting of the Association for
  Computational Linguistics, {ACL} 2020, Online, July 5-10, 2020}, pages
  7871--7880. Association for Computational Linguistics.

\bibitem[\protect\citename{Lewis \bgroup et al.\egroup }2020b]{lewis:2020a}
Lewis, P. S.~H., Perez, E., Piktus, A., Petroni, F., Karpukhin, V., Goyal, N.,
  K{\"{u}}ttler, H., Lewis, M., Yih, W., Rockt{\"{a}}schel, T., Riedel, S., and
  Kiela, D.
\newblock (2020b).
\newblock Retrieval-augmented generation for knowledge-intensive {NLP} tasks.
\newblock In {\em Advances in Neural Information Processing Systems 33: Annual
  Conference on Neural Information Processing Systems 2020, NeurIPS 2020,
  December 6-12, 2020, virtual}.

\bibitem[\protect\citename{Li \bgroup et al.\egroup }2021]{li2021ditch}
Li, H., Gao, T., Goenka, M., and Chen, D.
\newblock (2021).
\newblock Ditch the gold standard: Re-evaluating conversational question
  answering.
\newblock {\em arXiv preprint arXiv:2112.08812}.

\bibitem[\protect\citename{Lin}2004]{lin:2004}
Lin, C.-Y.
\newblock (2004).
\newblock {ROUGE}: A package for automatic evaluation of summaries.
\newblock In {\em Text Summarization Branches Out}, pages 74--81, Barcelona,
  Spain, July. Association for Computational Linguistics.

\bibitem[\protect\citename{Liu \bgroup et al.\egroup }2021a]{liu:2021}
Liu, Z., Zhou, K., Mao, J., and Wilson, M.~L.
\newblock (2021a).
\newblock {POSSCORE:} {A} simple yet effective evaluation of conversational
  search with part of speech labelling.
\newblock In Gianluca Demartini, et~al., editors, {\em 30th {ACM} International
  Conference on Information and Knowledge Management (CIKM'21)}, pages
  1119--1129. {ACM}.

\bibitem[\protect\citename{Liu \bgroup et al.\egroup }2021b]{liu2021b}
Liu, Z., Zhang, K., Xiong, C., Liu, Z., and Sun, M.
\newblock (2021b).
\newblock Openmatch: An open source library for neu-ir research.
\newblock In {\em {SIGIR} '21: The 44th International {ACM} {SIGIR} Conference
  on Research and Development in Information Retrieval, Virtual Event, Canada,
  July 11-15, 2021}, pages 2531--2535. {ACM}.

\bibitem[\protect\citename{Luan \bgroup et al.\egroup
  }2021]{DBLP:journals/tacl/LuanETC21}
Luan, Y., Eisenstein, J., Toutanova, K., and Collins, M.
\newblock (2021).
\newblock Sparse, dense, and attentional representations for text retrieval.
\newblock {\em Trans. Assoc. Comput. Linguistics}, 9:329--345.

\bibitem[\protect\citename{Potthast \bgroup et al.\egroup }2019]{potthast:2019}
Potthast, M., Gollub, T., Wiegmann, M., and Stein, B.
\newblock (2019).
\newblock {TIRA} integrated research architecture.
\newblock In Nicola Ferro et~al., editors, {\em Information Retrieval
  Evaluation in a Changing World - Lessons Learned from 20 Years of {CLEF}},
  volume~41 of {\em The Information Retrieval Series}, pages 123--160.
  Springer.

\bibitem[\protect\citename{Qiu \bgroup et al.\egroup
  }2021]{DBLP:conf/aaai/QiuHCJQ0HZ21}
Qiu, M., Huang, X., Chen, C., Ji, F., Qu, C., Wei, W., Huang, J., and Zhang, Y.
\newblock (2021).
\newblock Reinforced history backtracking for conversational question
  answering.
\newblock In {\em Thirty-Fifth {AAAI} Conference on Artificial Intelligence,
  {AAAI} 2021, Thirty-Third Conference on Innovative Applications of Artificial
  Intelligence, {IAAI} 2021, The Eleventh Symposium on Educational Advances in
  Artificial Intelligence, {EAAI} 2021, Virtual Event, February 2-9, 2021},
  pages 13718--13726. {AAAI} Press.

\bibitem[\protect\citename{Radford \bgroup et al.\egroup }2019]{radford:2019}
Radford, A., Wu, J., Child, R., Luan, D., Amodei, D., Sutskever, I., et~al.
\newblock (2019).
\newblock Language models are unsupervised multitask learners.
\newblock {\em OpenAI blog}, 1(8):9.

\bibitem[\protect\citename{Raffel \bgroup et al.\egroup }2020]{raffel:2019}
Raffel, C., Shazeer, N., Roberts, A., Lee, K., Narang, S., Matena, M., Zhou,
  Y., Li, W., and Liu, P.~J.
\newblock (2020).
\newblock Exploring the limits of transfer learning with a unified text-to-text
  transformer.
\newblock {\em J. Mach. Learn. Res.}, 21:140:1--140:67.

\bibitem[\protect\citename{Risch \bgroup et al.\egroup }2021]{risch:2021}
Risch, J., M{\"o}ller, T., Gutsch, J., and Pietsch, M.
\newblock (2021).
\newblock Semantic answer similarity for evaluating question answering models.
\newblock In {\em 3rd Workshop on Machine Reading for Question Answering},
  pages 149--157.

\bibitem[\protect\citename{Siblini \bgroup et al.\egroup
  }2021]{DBLP:conf/acl/SibliniSK20}
Siblini, W., Sayil, B., and Kessaci, Y.
\newblock (2021).
\newblock Towards a more robust evaluation for conversational question
  answering.
\newblock In {\em Proceedings of the 59th Annual Meeting of the Association for
  Computational Linguistics and the 11th International Joint Conference on
  Natural Language Processing, {ACL/IJCNLP} 2021, (Volume 2: Short Papers),
  Virtual Event, August 1-6, 2021}, pages 1028--1034. Association for
  Computational Linguistics.

\bibitem[\protect\citename{Vakulenko \bgroup et al.\egroup
  }2020]{vakulenko2020wrong}
Vakulenko, S., Longpre, S., Tu, Z., and Anantha, R.
\newblock (2020).
\newblock A wrong answer or a wrong question? an intricate relationship between
  question reformulation and answer selection in conversational question
  answering.
\newblock In {\em Proceedings of the 5th International Workshop on
  Search-Oriented Conversational AI (SCAI)}, pages 7--16.

\bibitem[\protect\citename{Vakulenko \bgroup et al.\egroup
  }2021]{DBLP:conf/wsdm/VakulenkoLTA21}
Vakulenko, S., Longpre, S., Tu, Z., and Anantha, R.
\newblock (2021).
\newblock Question rewriting for conversational question answering.
\newblock In Liane Lewin{-}Eytan, et~al., editors, {\em {WSDM} '21, The
  Fourteenth {ACM} International Conference on Web Search and Data Mining,
  Virtual Event, Israel, March 8-12, 2021}, pages 355--363. {ACM}.

\bibitem[\protect\citename{Voorhees}2003]{DBLP:conf/naacl/Voorhees03}
Voorhees, E.~M.
\newblock (2003).
\newblock Evaluating the evaluation: {A} case study using the {TREC} 2002
  question answering track.
\newblock In Marti~A. Hearst et~al., editors, {\em Human Language Technology
  Conference of the North American Chapter of the Association for Computational
  Linguistics, {HLT-NAACL} 2003, Edmonton, Canada, May 27 - June 1, 2003}. The
  Association for Computational Linguistics.

\bibitem[\protect\citename{Yang \bgroup et al.\egroup }2017]{yang:2017}
Yang, P., Fang, H., and Lin, J.
\newblock (2017).
\newblock Anserini: Enabling the use of lucene for information retrieval
  research.
\newblock In Noriko Kando, et~al., editors, {\em Proceedings of the 40th
  International {ACM} {SIGIR} Conference on Research and Development in
  Information Retrieval, Shinjuku, Tokyo, Japan, August 7-11, 2017}, pages
  1253--1256. {ACM}.

\bibitem[\protect\citename{Yu \bgroup et al.\egroup }2020]{yu:2020}
Yu, S., Liu, J., Yang, J., Xiong, C., Bennett, P.~N., Gao, J., and Liu, Z.
\newblock (2020).
\newblock Few-shot generative conversational query rewriting.
\newblock In Jimmy Huang, et~al., editors, {\em Proceedings of the 43rd
  International {ACM} {SIGIR} conference on research and development in
  Information Retrieval, {SIGIR} 2020, Virtual Event, China, July 25-30, 2020},
  pages 1933--1936. {ACM}.

\bibitem[\protect\citename{Zhang \bgroup et al.\egroup }2020a]{zhang:2020b}
Zhang, J., Zhao, Y., Saleh, M., and Liu, P.~J.
\newblock (2020a).
\newblock {PEGASUS:} pre-training with extracted gap-sentences for abstractive
  summarization.
\newblock In {\em Proceedings of the 37th International Conference on Machine
  Learning, {ICML} 2020, 13-18 July 2020, Virtual Event}, volume 119 of {\em
  Proceedings of Machine Learning Research}, pages 11328--11339. {PMLR}.

\bibitem[\protect\citename{Zhang \bgroup et al.\egroup }2020b]{zhang:2020}
Zhang, T., Kishore, V., Wu, F., Weinberger, K.~Q., and Artzi, Y.
\newblock (2020b).
\newblock Bertscore: Evaluating text generation with {BERT}.
\newblock In {\em 8th International Conference on Learning Representations
  ({ICLR}'20)}. OpenReview.net.

\bibitem[\protect\citename{Zobel and Rashidi}2020]{zobel:2020}
Zobel, J. and Rashidi, L.
\newblock (2020).
\newblock Corpus bootstrapping for assessment of the properties of
  effectiveness measures.
\newblock In Mathieu d'Aquin, et~al., editors, {\em {CIKM} '20: The 29th {ACM}
  International Conference on Information and Knowledge Management, Virtual
  Event, Ireland, October 19-23, 2020}, pages 1933--1952. {ACM}.

\end{thebibliography}


\end{document}